\documentclass{article}

\usepackage[english]{babel}

\usepackage[letterpaper,top=2cm,bottom=2cm,left=3cm,right=3cm,marginparwidth=1.75cm]{geometry}

\usepackage{amsmath}
\usepackage{graphicx}
\usepackage{float}
\usepackage{enumitem}
\usepackage{makecell}
\usepackage{authblk}
\usepackage{cite}
\usepackage[normalem]{ulem}
\usepackage[colorlinks=true, allcolors=blue]{hyperref}

\usepackage{booktabs}

\title{Accelerating kinetic plasma simulations with machine learning generated initial conditions}

\author[1]{Andrew T. Powis}
\author[1]{Doménica Corona Rivera}
\author[2]{Alexander Khrabry}
\author[1]{Igor D. Kaganovich}

\affil[1]{Princeton Plasma Physics Laboratory, Princeton, New Jersey 08540, USA}
\affil[2]{Princeton University, Princeton, New Jersey 08544, USA}

\begin{document}
\maketitle

\begin{abstract}
Computer aided engineering of multi-time-scale plasma systems which exhibit a quasi-steady state solution are challenging due to the large number of time steps required to reach convergence. Machine learning techniques combined with traditional first-principles simulations and high-performance computing offer many interesting pathways towards resolving this challenge. We consider acceleration of kinetic plasma simulations via machine learning generated initial conditions. The approach is demonstrated through modeling of capacitively coupled plasma discharges relevant to the microelectronics industry. Three models are trained on simulations across a parameter space of device driving frequency and operating pressure. The models incorporate elements of a multi-layer perceptron, principal component analysis, and convolutional neural networks to predict the final time-averaged profiles of ion-density and velocity distribution functions. These data-driven initial condition generators (ICGs) provide a mean speedup of 17.1x in convergence time, when measured using an offline procedure, or a 4.4x speedup with an online procedure, with convolutional neural networks leading to the best performance. The paper also outlines a workflow for continuous data-driven model improvement and simulation speedup, with the aim of generating sufficient data for full device digital twins.
\end{abstract}

\section{Introduction}
\label{sec:intro}

Following a transient startup phase, many plasma systems operate at a quasi-steady, as opposed to a true steady, equilibrium. Examples include systems driven by time-periodic forcing functions, those with persistent waves and instabilities, or chaotic and turbulent systems. Applications where such plasma states exist include electric propulsion devices \cite{choueiri2001plasma,boeuf2017tutorial}, semiconductor manufacturing equipment \cite{chabert2011physics,lieberman1994principles}, and fusion reactors which operate in a stead-state (as opposed to a pulsed) mode \cite{geiger2012aspects}. Further examples can be found across engineering, including fluid and structural mechanics \cite{billah1991resonance,bearman1984vortex}, chemical \cite{noyes1974oscillatory}, and electrical engineering \cite{kunjumuhammed2015electrical}. For computer-aided engineering (CAE), these systems are often modeled as initial value problems (IVPs), where the governing differential equations are numerically integrated until the system ``converges'' to a quasi-steady solution suitable for analysis. A major challenge for simulating plasma systems is the need to resolve both the fastest and slowest dynamics, which can lead to a prohibitively large number of time steps to reach convergence. This multi-time-scale challenge presents a significant bottleneck for CAE-driven design of plasma technologies, where rapid, cost-effective, and accurate exploration of parameter spaces is required.

The time to convergence of such systems is often dependent on the chosen initial condition (IC) for the IVP. Consider, for example, a simple linear first order ODE,

\begin{equation}
\label{eq:ex1}
    x' + x = \cos t + \sin t + c,
\end{equation}
where $c$ is a fixed constant. With IC $x(t=0)=x_0$, the solution to this ODE is,

\begin{equation}
\label{eq:ex2}
    x(t) = (x_0 - c) e^{-t} + \sin(t) + c,
\end{equation}
a time periodic function with a decaying initial component. At long times this solution tends to the quasi-steady solution: $x_{\infty} (t) = \lim_{t\rightarrow \infty} x(t) = \sin(t) + c$. With time averaged solution $\langle x_{\infty}(t) \rangle=c$. The error between the solution and quasi-steady solution at any time is,

\begin{equation}
    \epsilon (t) = x(t) - x_{\infty} (t) = (x_0 - c) e^{-t}.
    \label{eq:error_example}
\end{equation}

If the quasi-steady solution is sought with accuracy below some absolute tolerance $|\epsilon (t)| < \epsilon_c$, then Eq. \ref{eq:error_example} shows that the time to convergence $t_c \propto \ln{|x_0-c|}$. Therefore, the closer the IC $x_0$, is to the time average of the quasi-steady solution $c$, the faster the system will converge.

More generally, in a kinetic plasma system, we assume that $t_c \propto g(\mathbf{x}_0 - \mathbf{c})$, where $g(\mathbf{x})$ is a monotonic increasing function, and $\mathbf{x}_0$ and $\mathbf{c}$ denote the multi-dimensional initial plasma state and the time-averaged quasi-steady state solution, respectively. The governing equations are either the Vlasov-Poisson or Vlasov-Maxwell system, with appropriate collision terms and boundary conditions. This is a highly non-linear set of equations, preventing a simple procedure for obtaining the optimal plasma IC.

Data-driven approaches can offer a solution to this problem. More specifically, an initial condition generator (ICG) can be trained on prior simulation data to select $\mathbf{x}_0 \approx \mathbf{c}$, with the goal of reducing the number of steps to convergence. For high-dimensional, multi-scale, or multi-physics problems, the amount of data required to learn an accurate mapping from design parameters to optimal ICs may be prohibitively large. This challenge can be mitigated by reducing the dimensionality of the IC, focusing on large-scale and long-time features of the system. Once a suitable low-dimensional IC is identified, the full system state can be recovered from the high-fidelity simulation, with accelerated convergence provided by the ICG.

Machine learning (ML) techniques are starting to see broader adoption within the plasma physics community \cite{anirudh20232022} as well as being specifically applied to low-temperature plasmas \cite{faraji2025machine,trieschmann2023machine,bonzanini2023foundations}. The methodology described here is considered a type of hybrid machine-learning combined with traditional high-performance computing (ML+HPC) \cite{von2020combining,wang2022hybrid,procopio2023combined}, whereby the merits of both techniques are leveraged to accelerate simulations while retaining the same level of accuracy of first principles simulations. Examples of data-generated ICs for accelerating simulations include work by Hajgat\'o et al. \cite{hajgato2019accelerating} who leveraged a convolutional neural network to realize a 13\% speedup in fluid dynamics simulations for the optimization of aerodynamic design. Barros et al. \cite{barros2024using} relied on an automatic hyperparameter optimization tool to train a recurrent neural network, achieving an 8\% speedup in solar wind forecasting models. Despite the simplicity of this concept, we build on previous work to demonstrate that for highly multi-scale plasma systems an order of magnitude improvement in convergence time can be achieved.

While the outlined procedure has the ML component out-of-the-loop, there exists a natural workflow for continuous model improvement which integrates ML more directly into the parameter space exploration. A well trained ICG can be used to accelerate simulations, which can in turn be used to collect more data to improve performance of the ICG. Iteration of this process may lead to sufficient data collection to produce a highly accurate reduced order model (ROM), or digital twin (DT), of the system \cite{sharma2022digital}. It should be emphasized that this approach is most practical when deep exploration of a parameter space is required, either for the purposes of design optimization or real time control \cite{lin2025digital,behrendt2025real,haddod2021intelligent}. As such, the approach described here may prove less beneficial for scientific discovery simulations, which require exploration of completely unexplored regions of a parameter space.

\subsection{Motivating example: capacitively coupled plasma discharges for microelectronics manufacturing}
\label{sec:intro_ccp}

\begin{figure}
    \centering
    \includegraphics[width=0.7\linewidth]{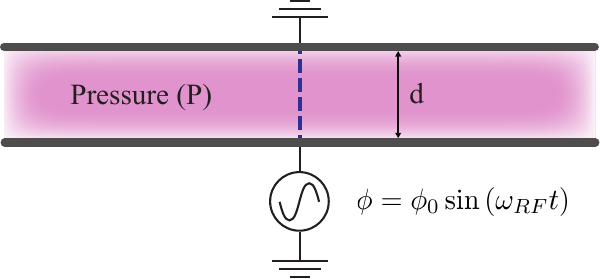}
    \caption{Two-dimensional schematic of an argon  capacitively coupled plasma discharge with gap width $d$, gas pressure $P$, driving voltage amplitude $\phi_0$ and angular frequency $\omega_{RF}=2 \pi F$. The blue dashed line indicates the 1D spatial domain modeled by 1D-3V PIC simulations in this work.}
    \label{fig:ccp_intro}
\end{figure}

To demonstrate the advantages of a data-driven ICG, we consider the application of low-temperature plasmas to the etching of silicon wafers. Plasmas are used in nearly 50\% of all steps in the complex manufacturing process of turning raw silicon into microcircuitry \cite{kanarik2020inside}, including the generation of extreme ultraviolet light for photolithography \cite{kazazis2024extreme} as well as chemical doping, cleaning, and etching of silicon surfaces \cite{national2020plasma}. Capacitively coupled plasma (CCP) discharges are commonly used in the dry etching of silicon wafers \cite{lieberman1994principles,chabert2011physics} (see the schematic in Fig. \ref{fig:ccp_intro}). A time-periodic waveform between parallel plates (with frequency $\omega_{RF}=2 \pi F$) drives large scale oscillations of plasma electrons. At high pressures, this leads to ohmic heating, whereas at low pressures, stochastic heating processes dominate. The heated electrons enable chemical reactions which sustain the plasma and drive reactions at the wafer surface. Heavier plasma ions (with plasma frequency $\omega_{pi} \ll \omega_{RF}$) perceive the radio-frequency (RF) sheath as a strong constant electric field, which accelerate these particles into the silicon surface.

As traditional approaches for shrinking transistor size offer diminishing returns, the semiconductor industry is exploring more innovative approaches to sustaining Moore's law \cite{graves2024science}. A prominent example includes complex three-dimensional microcircuit geometries, which can effectively stack circuit components vertically at the nanometer scale \cite{kim2017evolution}. High-aspect ratio etching of deep channels into the wafer is an essential step to realizing these structures, requiring unprecedented control over ion kinetics at the plasma-surface interface \cite{adamovich20222022}. This has pushed the operating pressures of CCPs into the low or even sub-mTorr range \cite{liu2011collisionless,kawamura2006stochastic,sharma2022investigating,patil2022electron,zhang2022ohmic}, where improved control over the ion velocity distribution function (IVDF) can be achieved. Resolving these anisotropic IVDFs, as well as electron stochastic heating and sheath-plasma instabilities \cite{wilczek2020electron} necessitates a kinetic treatment of the system. High resolution first principles kinetic plasma simulations therefore play an essential role in understanding and optimizing CCP discharges to maximize wafer yield and throughput.

Beyond the need for high-dimensional kinetic resolution, high-fidelity simulations of CCPs are also challenging due to the large mass ratio between electrons and ions, leading to the aforementioned difficulty of multi-time-scale modeling. A requirement for numerical and physical accuracy is the resolution of the short time scale electron plasma period ($\omega_{pe}$) \cite{wilczek2020electron,birdsall2018plasma}, whereas the system converges on the much longer time scale of ion transport. This generally results in a minimum $10^6$ and up to $10^8$ time steps to reach quasi-steady convergence. The most common technique used for modeling CCPs is the electrostatic particle-in-cell (PIC) algorithm with Monte Carlo methods applied to model collisions between charged species and the neutral gas \cite{hockney2021computer,birdsall2018plasma,tajima2018computational,vahedi1995monte,donko2011particle}. The grid and particles in the PIC algorithm are highly amenable to parallelization and acceleration \cite{burau2010picongpu,juhasz2021efficient}, however the tightly coupled set of equations is advanced serially in time, limiting opportunities for parallelization and therefore further acceleration.

In general, RF-CCPs have the following ordering of time scales, $\omega_{pe} \gg \omega_{RF} \gg \omega_{pi}$, meaning that the perturbation of the IVDF about the mean solution is small and can be considered representative of a time-averaged steady-state solution for the CCP (the $\mathbf{c}$ from Section \ref{sec:intro}). The goal is therefore to create an ICG which can either predict the time-averaged IVDF (which we will simply refer to as $f$, without the species subscript), or moments of this distribution function. The implied hypothesis is that the lighter electrons will rapidly adapt to this initial state and thereby accelerate convergence to the final quasi-steady solution.

The paper proceeds as follows. Section \ref{sec:method} outlines the PIC code used to create the data and various machine learning methods used to train the ICG. The section also discusses how we measure convergence of our simulations and set a baseline for the proposed ML performance improvements. Section \ref{sec:acceleration} will then demonstrate that the proposed approach does indeed improve performance by over an order of magnitude. Section \ref{sec:digital_twin} will outline how this procedure can be integrated into a workflow for generating reduced order models or digital twins of plasma reactors. Section \ref{sec:conclusion} will then conclude the findings of this research.

\section{Methodology}
\label{sec:method}

\subsection{Capacitively coupled plasma discharge simulations}

To demonstrate the effectiveness of a data-driven ICG for accelerating CCP simulations, we consider a reduced dimension, 1D-3V CCP configuration (see the dashed line in Fig. \ref{fig:ccp_intro}). Parameters which strongly influence the operating conditions of the discharge include the ion and gas species, gas pressure ($P$), electrode gap ($d$) and, assuming a simple sinusoidal waveform, the voltage amplitude ($\phi_0$) and frequency ($F$). For this demonstration, only $P$ and $F$ are allowed to vary with the species fixed to argon, $d = 5cm$, and $\phi_0=300V$. All simulations satisfy the appropriate criteria for numerical convergence, including resolution of the electron plasma Debye length and electron plasma period, and have sufficient particles-per-cell for statistical accuracy \cite{jubin2024numerical}. The state of the system is initialized with a uniform Maxwellian distribution function with density $n_0$, electron temperature $T_{e0}$, and ion temperature $T_{i0}$.

Simulations are completed using the 1D-3V Python PIC code \texttt{mini-pic} \cite{powis2024accuracy}, a parallel and accelerated code designed for prototyping new algorithms. The code has been well benchmarked on previous CCP simulations\cite{turner2013simulation,powis2024accuracy}. While the code can operate on GPUs, for 1D simulations, best performance was observed when leveraging multiple CPU cores through \texttt{mpi4py} \cite{dalcin2021mpi4py}.

250 1D-3V PIC simulations of CCP discharges were run across the range of conditions listed in Table \ref{tab:ccp1d_params}. These simulations were completed in 2 days using 17 nodes on the Princeton Stellar cluster, where each node consists of four 2.9 GHz Intel Cascade Lake CPUs, each with 24 cores (for a total of 1,632 cores). After reaching convergence, plasma properties were averaged over 100 RF periods to provide final $f$ profiles (time-averaged IVDFs).

\begin{table}[H]
\centering
  \caption{Physical and numerical parameters for the set of 1D-3V CCP simulations used to train the ICG models.}
  \label{tab:ccp1d_params}
  \begin{tabular}{r|c}
    \toprule
    Parameter&Value\\
    \midrule
    Ion species & argon\\
    Gas pressure range ($P$) & 2-30 mTorr\\
    Gap width ($d$) & 5 cm\\
    Driving frequency range ($F$) & 6.78-50.85 MHz\\
    Driving amplitude ($\phi_0$) & 300 V\\
    Initial electron temperature ($T_{e0}$) & 1.5 eV\\
    Initial ion temperature ($T_{i0}$) & 300 K\\
    Number of simulations & 250\\
    Number of cells & 64-1,408\\
    Number of steps & 3-31 million\\
    Step size & 7.12-236 ps\\
    Simulated time & 189-1,688 $\mu s$\\
    Mean particles-per-cell & $\sim 300$ per species\\
  \bottomrule
\end{tabular}
\end{table}

\subsection{Definition for particle-in-cell simulation convergence}
\label{sec:convergence_def}

Due to the noisy and statistical nature of PIC simulations, defining a rigorous measure of simulation convergence is challenging. A useful decision algorithm for stopping a practical simulation should ideally be robust, cheap to compute, and performed online.

Attempting to satisfy these criterion, we consider the 1D steady state ion continuity equation,

\begin{equation}
    \frac{\partial}{\partial x}\int v f_{i} dv \equiv \frac{\partial \Gamma_i}{\partial x} = S_{+}(x),
\end{equation}
where $\Gamma_i(x)$ is the ion flux density and $S_+(x)$ represents the ionization source term. Notably, the kinetic behavior of the plasma is implicitly embedded within the source term, since $S_+(x)$ depends on the electron distribution function. As a result, the equation is capable of capturing kinetic effects without the need for expensive diagnostics of the full kinetic distribution functions.

Integrating both sides yields,

\begin{equation}
    \label{eq:conv0}
    \Gamma_i(x) = \int_0^x S_{+}(x) dx + A \equiv \Lambda(x).
\end{equation}

\begin{figure}[H]
    \centering
    \includegraphics[width=0.7\linewidth]{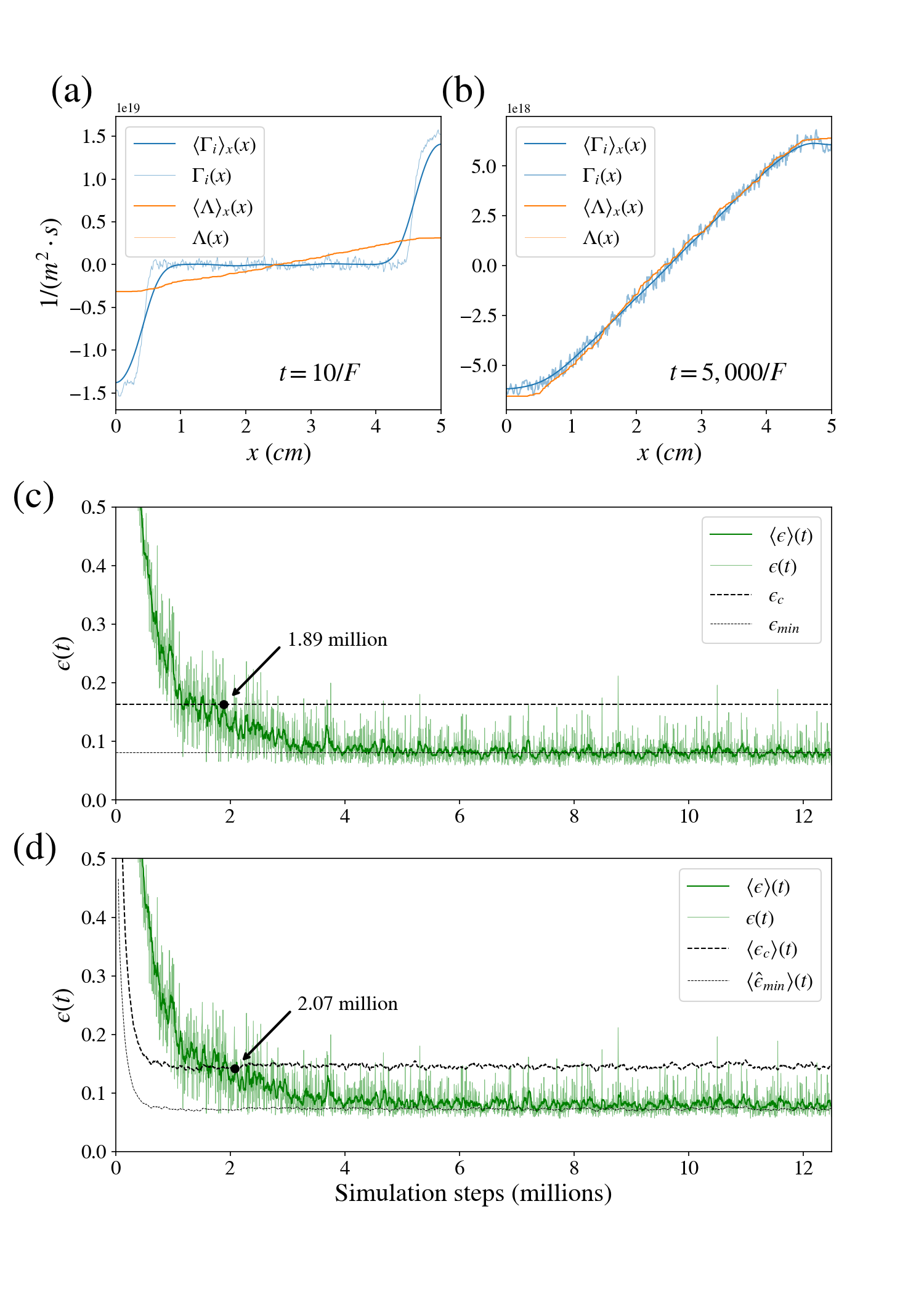}
    \caption{(a) Plots of $\Gamma_i$ and $\Lambda$ early in the simulation (at 10 RF periods) and, (b) after convergence has been achieved (at 5,000 RF periods) for 1D-3V PIC simulations of an argon CCP operating at $F=27.12$ $MHz$ and $P = 20$ $mTorr$. Thicker lines correspond to functions smoothed over space with a Gaussian filter. (c) Evolution of $\epsilon(t)$ with time including the point of convergence measured by the \textit{offline} procedure. The thicker line corresponds to the same data smoothed with a centered uniform filter. (d) Evolution of $\epsilon(t)$ and $\langle \epsilon_c \rangle(t)$ with time including the point of convergence measured by the \textit{online} procedure. Temporal smoothing is realized with a backward uniform filter.}
    \label{fig:conv_example}
\end{figure}

The integration constant $A$ is selected such that the mean of $\Gamma_i(x)$ and $\Lambda(x)$ over $x$ is zero. To reduce the effects of numerical noise, $\Gamma_i$ and $\Lambda$ values are time-averaged over a fixed size sliding window $T_a=a/F$. At the end of each window the data can be used to test for convergence and make a decision on stopping or continuing the simulation, or be stored for later (offline) analysis. A value of $a=5$ was found to be a good compromise for reducing noise, while providing sufficient temporal granularity for useful convergence testing. For CCP settings $\{$$F=27.12$ $MHz$, $P = 20$ $mTorr$$\}$, Fig. \ref{fig:conv_example}a and b show examples of time-averaged $\Gamma_i$ and $\Lambda$ early in the simulation and after convergence.



At each time step, the level of convergence is measured via a symmetric normalized Euclidean norm of the difference between both sides of Eq. \ref{eq:conv0},

\begin{equation}
    \label{eq:conv1}
    \epsilon(t) = \frac{\left\| \Gamma_i (x,t) - \Lambda(x,t) \right\|_2}{ \frac{1}{2} \left\| \Gamma_i (x,t) \right\|_2 + \frac{1}{2} \left\|\Lambda(x,t) \right\|_2}.
\end{equation}

Figure \ref{fig:conv_example}c shows a plot of $\epsilon(t)$ vs time for the same example simulation. This clearly demonstrates that the noisy nature of PIC (especially for the $\Gamma_i(x)$ data) means that $\epsilon(t)$ never truly converges to zero, but rather oscillates around some minimum value, which we will call $\epsilon_{min}$.



As stated, an effective test for convergence should be \textit{online}, that is, the algorithm should only rely on data from the current and previous time-steps. However, to obtain a more accurate measure of convergence for the purposes of performance testing our ICG, we will also consider an \textit{offline} procedure.

\subsubsection{Offline test for convergence}
\label{sec:offline_conv}

All simulations are run to a total time $T_s=5,000/F$, where convergence can be confidently guaranteed. $\epsilon_{min}$ is then computed as the mean value of $\epsilon(t)$ for $t \in [\frac{3}{4}T_s,T_s]$. The level for convergence is set at two times this minimum noise level, such that $\epsilon_c = 2 \epsilon_{min}$. To further reduce the effects of noise and outliers, $\epsilon(t)$ is smoothed via a centered uniform filter which returns the mean across 11 time steps ($\langle \epsilon \rangle (t)$ in the figure). Convergence is then defined as the time at which $\langle \epsilon \rangle (t) < \epsilon_c$, and remains so until the simulation completes. Figure \ref{fig:conv_example}c shows the temporal evolution of these quantities and the point of convergence for the example simulation.

\subsubsection{Online test for convergence}

An online test for convergence obviously cannot rely on future information regarding $\epsilon (t)$, which complicates the procedure for measuring the minimum simulation noise level $\epsilon_{min}$. To overcome this, we return to the spatial information contained within $\Gamma_i(x)$ and $\Lambda (x)$. First, spatially smoothed profiles of each function are obtained via a centered Gaussian filter with standard deviation equal to 25 cells. The underlying PIC noise minimum is then estimated as the sum of the norms of the residuals from each function, such that,

\begin{equation}
    \hat{\epsilon}_{min} (t) = \frac{\left\| \Gamma_i (x,t) - \langle \Gamma_i \rangle_x (x,t) \right\|_2}{\left\| \Gamma_i (x,t) \right\|_2} + \frac{\left\| \Lambda (x,t) - \langle \Lambda \rangle_x (x,t) \right\|_2}{\left\| \Lambda (x,t) \right\|_2}.
\end{equation}

Where the estimate, $\hat{\epsilon}_{min} (t)$, is now a time-evolving quantity. The convergence limit is therefore also evolving in time as $\epsilon_c(t)= 2 \hat{\epsilon}_{min} (t)$. To further reduce the effects of temporal noise, both $\epsilon(t)$ and and $\epsilon_c(t)$ are temporally averaged over the proceeding 10 periods (giving $\langle \epsilon \rangle (t)$ and $\langle \epsilon_c \rangle(t)$ respectively). 

Another challenge associated with an online test for convergence is in deciding whether a dip in the convergence signal below the convergence criterion is permanent or temporary. To address this, we assume that if $\langle \epsilon \rangle (t) < \langle \epsilon_c \rangle(t)$ for 25 consecutive periods (in this case $25a=125/F$) then convergence has been achieved. Figure \ref{fig:conv_example}d shows the temporal evolution of these quantities and the point of online convergence for the example simulation.

The parameters chosen for this online procedure were selected through testing of the offline and online procedure over the 250 CCP simulations completed for this work. While these parameters proved to be robust over the range of CCP input parameters, we recommend that care be taken when adopting this procedure and tuning may be required for different problems and numerical settings.

Since the online procedures incorporates a waiting time to ensure that convergence has truly been achieved, it provides more conservative estimates of convergence time (i.e longer times) that those from the offline procedure. While this behavior is preferable, an overestimate of convergence time will negatively influence measurements of the performance improvements provided by the ICGs tested in this work (see Section \ref{sec:results_summary} below). Therefore this online procedure leaves some performance on the table, and we encourage future research into more advanced procedures to test for PIC convergence which may allow for the full acceleration afforded by ICGs (or other techniques) to be realized.

\subsection{Machine learning techniques for initial condition generation}

The task is to produce an initial condition generator (ICG) which predicts the time-averaged IVDF $f (x,\mathbf{v})$ after the system has reached quasi-steady state. Since the simulations are one-dimensional in space, and collisions work to isotropize the distribution function, it is reasonable to assume that the velocity distributions in the two directions parallel to the CCP electrodes (i.e. directions which are not spatially resolved) will remain nearly isotropic during evolution. We therefore further restrict our machine learning model to consider only the 1D-1V $f (x,v)$ distribution (where $v\equiv v_x$. We also consider an even simpler model for predicting only the 0th moment of the IVDF (ion density) defined as $n (x) = \int f (x,v) dv$.

In summary, we consider learning three mappings:

\begin{align}
    \Theta_1:\{F,P\} & \rightarrow n(x)\\
    \Theta_{2,3}:\{F,P\} & \rightarrow f(x,v)    
\end{align}

The output of these mappings have vastly different dimensions and therefore different machine learning techniques will likely prove advantageous for each case. For the $\Theta_1$ mapping we directly apply a classic Multilayer Perceptron (MLP) architecture. For the second mapping we consider two approaches; a Principal Component Analysis (PCA) decomposition of the data in combination with an MLP ($\Theta_2$), as well as a Convolution Neural Network (CNN) architecture ($\Theta_3$). The three procedures are outlined below.

\subsubsection{An MLP for predicting ion density}
\label{sec:meth_dnn}

To train the $\Theta_1$ mapping, ion density from the PIC particle data is interpolated onto a 1D spatial grid, such that $n_{j}=n(x_j)$, where each $x_j$ represents a grid node. Due to the inherent symmetry of the time-averaged ion density profiles about the domain midpoint, the size of each profile is reduced by half along the symmetry line (at $x=d/2$). Since the profiles are monotonic and relatively smooth, the spatial resolution is down-sampled to 16 uniformly spaced points. Additionally, the ion density in CCPs exhibits a power-law dependence on input parameters $F$ and $P$ \cite{fu2020similarity}, meaning that transforming the data into a logarithmic scale (using a softplus function), will improve learning performance. Of the 250 simulations available, 180 were used for training, with 30 reserved for validation and hyperparameter tuning, and 40 for final testing.

As a universal approximator for non-linear functions, a simple MLP is trained to learn this mapping, implemented via the \texttt{PyTorch} framework \cite{paszke2019pytorch,stevens2020deep}. The model incorporates 3 hidden fully connected layers, each with 64 neurons and ReLU activation functions. A Mean Squared Error (MSE) loss function combined with a softplus provided good performance and ensured positivity of generated density outputs. The Adam optimizer with a learning rate of 0.002 provided good convergence within 500 epochs. Full details of the network and optimized hyperparameters are provided in Table \ref{tab:dnn_params}. Training was extremely cheap, and could be completed on a laptop CPU in under 10 seconds

\begin{table}[H]
    \centering
  \caption{Data and hyperparameters used to train the MLP based ICG.}
  \label{tab:dnn_params}
  \begin{tabular}{r|c}
    \toprule
    Parameter&Value\\
    \midrule
    Number of train samples & 180 (72\%)\\
    Number of validation samples & 30 (12\%)\\
    Number of test samples & 40 (16\%)\\
    Training data type & $n_j$, $j \in [1,16]$ \\
    \midrule
    Optimizer & Adam\\
    Loss function & MSE \& softplus\\
    Activation function & ReLU\\
    Number of epochs & 500\\
    Learning rate & 0.002\\
    Input parameters & $\{F,P\}$\\
    Number of hidden layers & 3\\
    Hidden layer widths & $[64,64,64]$\\
    Output data type & Same as training\\
  \bottomrule
\end{tabular}
\end{table}

\subsubsection{A PCA \& MLP for predicting IVDF}
\label{sec:meth_pca_dnn}

To learn the $\Theta_2$ mapping, PIC particle data is interpolated onto a 1D-1V phase space grid, such that $f_{j,k}=f(x_j,v_{k})$, where each $x_j$ and $v_{k}$ represent nodes of the phase space grid. $x_j$ spans the 5cm gap with 400 points and $v_k$ spans $-\frac{1}{2}v_{i,RF}$ to $\frac{1}{2}v_{i,RF}$ with 512 points, where $v_{i,RF}=\sqrt{2e \phi_0 /m_i}$ is the velocity of an ion after acceleration through the RF potential amplitude. Following Section \ref{sec:meth_dnn} the data is also transformed into a logarithmic scale. To improve model accuracy, only simulations which produced non-zero final density were considered, with 195 used for training and 39 reserved for validation and testing.

As can be seen in Fig. \ref{fig:converge_test_fc} below, the IVDF profiles have a central low energy high density region and two tails associated with ion acceleration through the RF sheaths. A PCA decomposition \cite{pearson1901liii} is leveraged to reduce the dimensionality of the data for improved training with an MLP. PCA seeks a decomposition that exposes the maximum amount of variance across the dataset (i.e. across all simulations) for a given number of components. The decomposition is computed on the training data only and whitening was used to better normalize each component. A threshold of $91\%$ explained variance is set, requiring 27 PCA components.

\begin{figure}[H]
    \centering
    \includegraphics[width=0.9\linewidth]{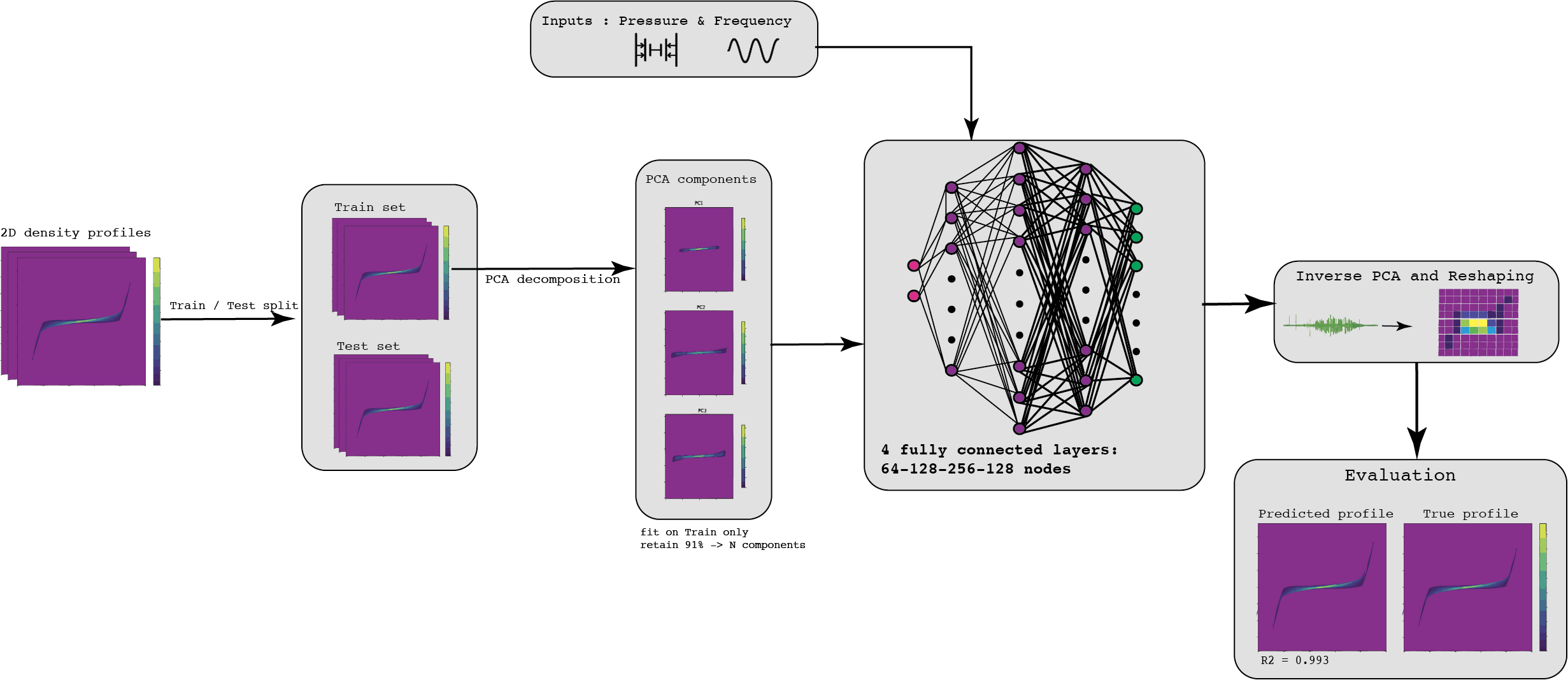}
    \caption{Workflow and model structure for the PCA+MLP training and resulting initial condition generator (ICG). The data is initially split into train and test samples. Dimensionality is reduced via a 27 component PCA decomposition, retaining 91\% of explained variance. The eigenvalues of these components for a given CCP frequency and pressure are then learned via an MLP with 4 hidden layers. Initial conditions are generated by first predicting the PCA eigenvalues from the MLP and then reconstructing the profiles.}
    \label{fig:pca_fnn_workflow}
\end{figure}

The respective eigenvalues of each PCA mode are then used to train the MLP, with the architecture and optimal parameters shown in Table \ref{tab:pca_params}. A workflow for the training procedure is outlined in Fig. \ref{fig:pca_fnn_workflow}. Total training time was 90 seconds on a single NVIDIA A100 GPU, a speedup of 48x compared to training on a single a single core of a 2.9 GHz Intel Cascade Lake CPU.

\begin{table}[H]
    \centering
  \caption{Data and hyperparameters for training the PCA+MLP based ICG.}
  \label{tab:pca_params}
  \begin{tabular}{r|c}
    \toprule
    Parameter&Value\\
    \midrule
    Number of train samples & 195 (83\%)\\
    Number of test samples & 39 (17\%)\\
    Training data type & \makecell{$f_{j,k}$ \\ $j \in [1,400]$\\$k \in [1,512]$}\\
    Explained variance & 91\%\\
    PCA components & 27\\
    \midrule
    Optimizer & Adam\\
    Loss function & MSE\\
    Activation function & ReLU\\
    Number of epochs & 3,000\\
    Learning rate & 2.2$\times$10$^{-4}$\\
    Input parameters & $\{F,P\}$\\
    Number of hidden layers & 4\\
    Hidden layer widths & $[64,128,256,128]$\\
    Output data type & Same as training\\
  \bottomrule
\end{tabular}
\end{table}

Generation of IVDF profiles is achieved by outputting PCA eigenvalues from the trained MLP. Full reconstruction of the IVDF can then be obtained by expanding the PCA components with their respective eigenvalues.

\subsubsection{A CNN for predicting IVDF}
\label{sec:meth_cnn}

For the CNN model, data is prepared by mapping the IVDF onto a 2D grid with the same limits as from Section \ref{sec:meth_pca_dnn}, although with 128 points in both the $x$ and $v$ directions. A logarithmic scale was not found to be necessary for improving performance of this model. Similarly, 195 simulations (with non-zero density) were used for training and 39 reserved for validation and testing.

\begin{figure}[H]
    \centering
    \includegraphics[width=0.7\linewidth]{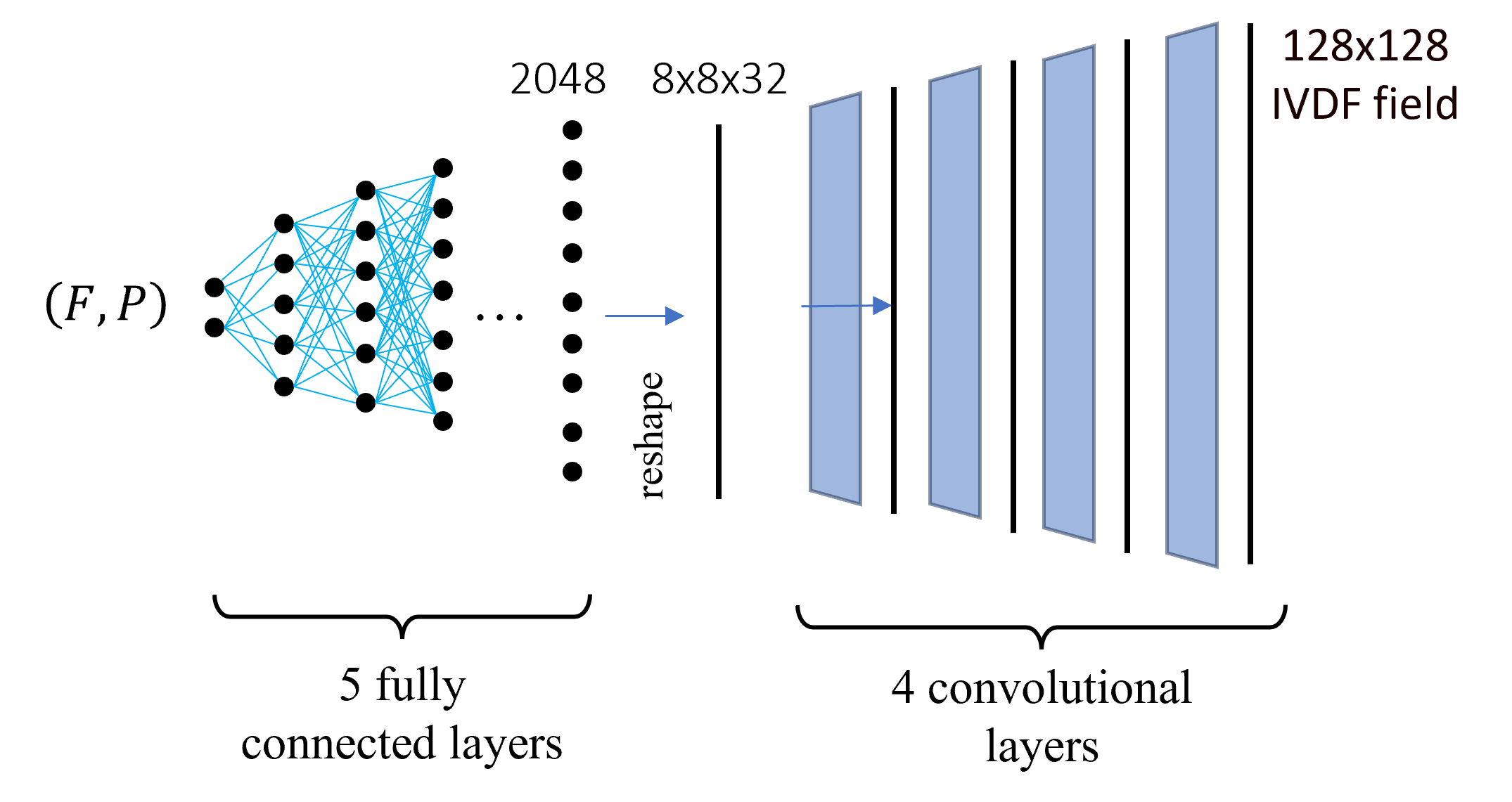}
    \caption{Model structure of the CNN trained to produce an initial condition generator (ICG) for the IVDF. The model includes 6 fully connected layers, followed by reshaping and then 4 convolutional layers.}
    \label{fig:cnn_model}
\end{figure}

A convolutional neural network leverages the structure of the data in a different way than a PCA decomposition. The model directly includes layers which ``learn'' the spatial structure of the data through convolutional or deconvolutional operations (in practice interpolation). Although the data here is two-dimensional, the convolutional operations include a third-dimension, known as ``channels'', which can learn and capture different features of the data. Figure \ref{fig:cnn_model} shows the structure of the CNN model, which first includes a 6 layered MLP to upsample the CCP input parameters before reshaping the data for the deconvolutional layers. The full details of the model are provided in Table \ref{tab:cnn_params}. Total training time was 20 seconds on a single NVIDIA A100 GPU.

\begin{table}[H]
    \centering
  \caption{Data and hyperparameters for training the CNN based ICG.}
  \label{tab:cnn_params}
  \begin{tabular}{r|c}
    \toprule
    Parameter&Value\\
    \midrule
    Number of train samples & 195 (83\%)\\
    Number of test samples & 39 (17\%)\\
    Training data type & \makecell{$f_{j,k}$ \\ $j,k \in [1,128]$}\\
    \midrule
    Optimizer & Adam\\
    Loss function & MSE\\
    Activation function & CELU\\
    Number of epochs & 1,000\\
    Learning rate & 0.001\\
    Input parameters & $\{F,P\}$\\
    Number of MLP hidden layers & 6\\
    Hidden layer widths & $[10,64,256,1024,2048,2048]$\\
    Number of CNN hidden layers & 4\\
    CNN layer kernel size & 4 \\
    CNN layer stride & 2 \\
    CNN layer padding & 0 \\
    CNN layer output channels & [16,8,4,1] \\
    Output data type & Same as training\\
  \bottomrule
\end{tabular}
\end{table}

\subsection{Establishing a baseline and measuring simulation speedup}

Establishing an appropriate baseline for machine learning enabled improvements of simulation performance is an essential part of evaluating the merits of any technique \cite{mcgreivy2024weak}. Since we consider approaches to improve the initialization of CCP simulations, this raises the question; how do subject matter experts usually initialize their simulations? In the author's experience, most modelers will initialize simulations with a uniform density Maxwellian distribution of electrons and ions, similar to those used for data collection in this paper. Choosing a practical initial density $n$ and species temperatures $T_{e0}$, $T_{i0}$ can be determined from literature or the modelers previous experience with similar simulations. To define a more rigorous baseline we rely on a 0D non-uniform CCP global model \cite{chabert2011physics} to predict $n$ and $T_{e0}$, with ion temperature always initialized as $300$ $K$. Full details of the global model and solution algorithm are provided in Appendix B.

The potential speedup provided by each of the ICGs when compared to this baseline is measured on the reserved test data which was not used during the training of the models. Each of the CCP simulations were run with six different initial conditions
\begin{enumerate}
    \item A uniform density determined by the global model.
    \item The exact time-averaged density profile obtained from converged simulations.
    \item The exact time-averaged IVDF profile obtained from converged simulations.
    \item The density profile (with Maxwellian velocity distribution) determined by $\Theta_1$.
    \item The IVDF profile determined by $\Theta_2$ (PCA+MLP) and electron density profile determined by integrating the IVDF.
    \item The IVDF profile determined by $\Theta_3$ (CNN) and electron density profile determined by integrating the IVDF.
\end{enumerate}
PIC initial conditions are created via rejection sampling over the respective $n$ or $f$ profiles.

At a single operating condition, simulation speedup (or slowdown) is measured as the ratio of the number of steps required to reach convergence (either by the online or offline procedure). Speedup is always measured against the baseline, that is, the number of steps required for a simulation to reach convergence when initialized with a uniform density predicted by the global model. Across the entire dataset, speedup is measured as the mean speedup for each ICG type.

We stress that the final converged state of each CCP simulation is identical, within numerical tolerance, regardless of the initial condition chosen. We therefore consider the described approach to be a strong baseline for measuring ML model performance.

\section{Results: Accelerating convergence}
\label{sec:acceleration}

\subsection{Convergence time with ideal initial conditions}
\label{sec:test_hypothesis}

\begin{figure}[H]
    \centering
    \includegraphics[width=0.6\linewidth]{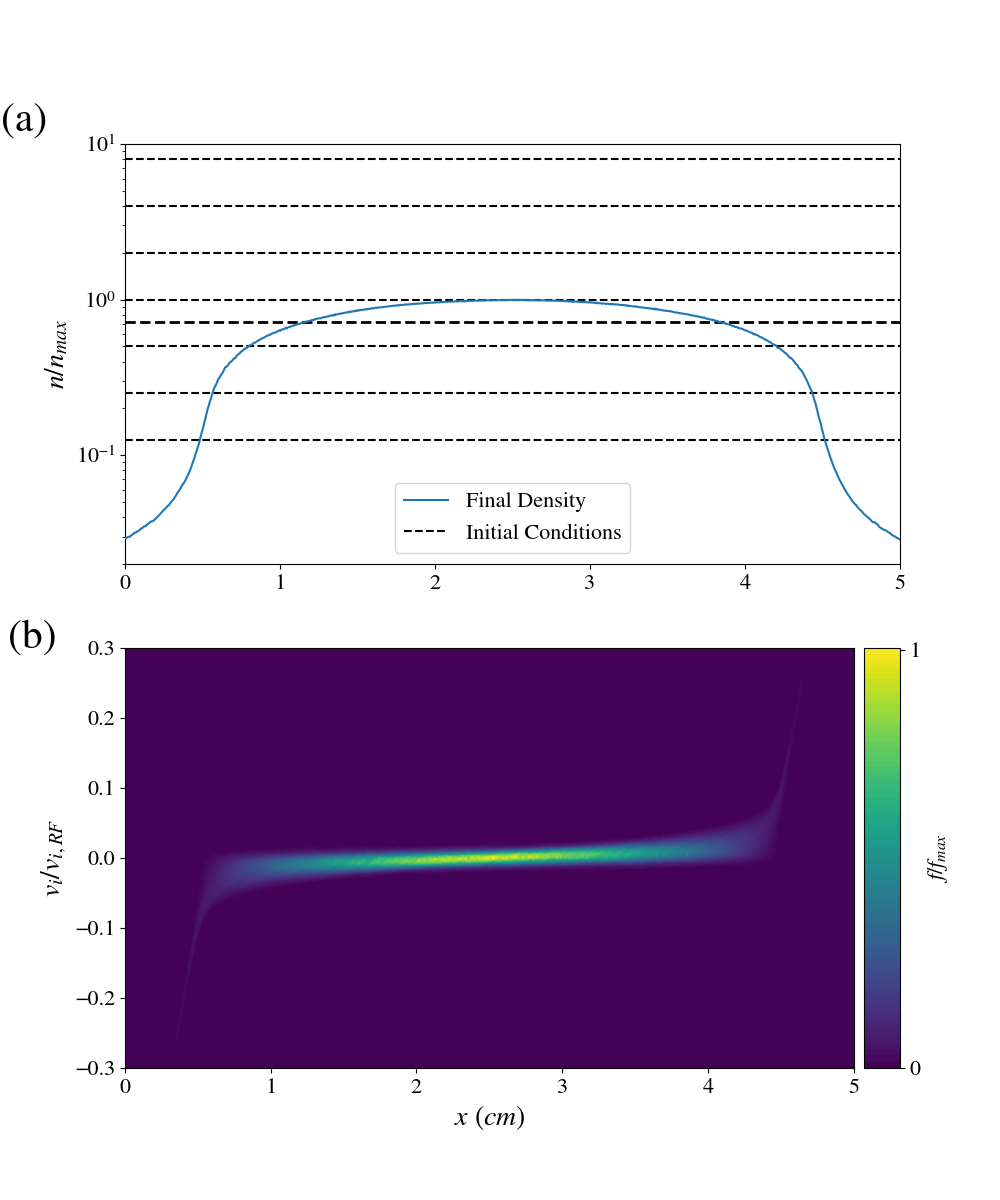}
    \caption{(a) Converged time-averaged ion density profile (blue line) of an argon CCP discharge operating at 27.12 MHz with 20 mTorr gas pressure and a range of uniform initial conditions (dashed lines). The thicker dashed line corresponds to the density predicted by the global model. (b) Converged time-average ion velocity distribution function (IVDF) for the same discharge parameters.}
    \label{fig:converge_test_fc}
\end{figure}

To verify that improved initial conditions can accelerate kinetic RF-CCP simulations we begin by considering a single point in the CCP parameter space $\{$$F=27.12$ $MHz$, $P = 20$ $mTorr$$\}$. Figure \ref{fig:converge_test_fc}a shows the resulting time-averaged steady-state ion density profile $n$ (blue line), and Fig. \ref{fig:converge_test_fc}b the time-averaged steady-state IVDF $f$. 

The simulations are run to convergence with various initial conditions:

\begin{enumerate}
    \item A series of uniform density profiles, as shown by the dashed lines in Fig. \ref{fig:converge_test_fc}a.
    \item The exact time-averaged ion density profile, as shown by the blue solid line in Fig. \ref{fig:converge_test_fc}a.
    \item The exact time-averaged IVDF, as shown in Fig. \ref{fig:converge_test_fc}b.
\end{enumerate}

Electrons are initialized with the same density profile as ions, either directly from $n$ or integrated from $f$ and with an isotropic Maxwellian velocity distribution. Unspecified components of the IVDF are also initialized in the same way (i.e. in the $v_y,v_z$, direction) with Maxwellian distributions using the characteristic temperatures given in Table \ref{tab:ccp1d_params}.

\begin{figure}[H]
    \centering
    \includegraphics[width=0.8\linewidth]{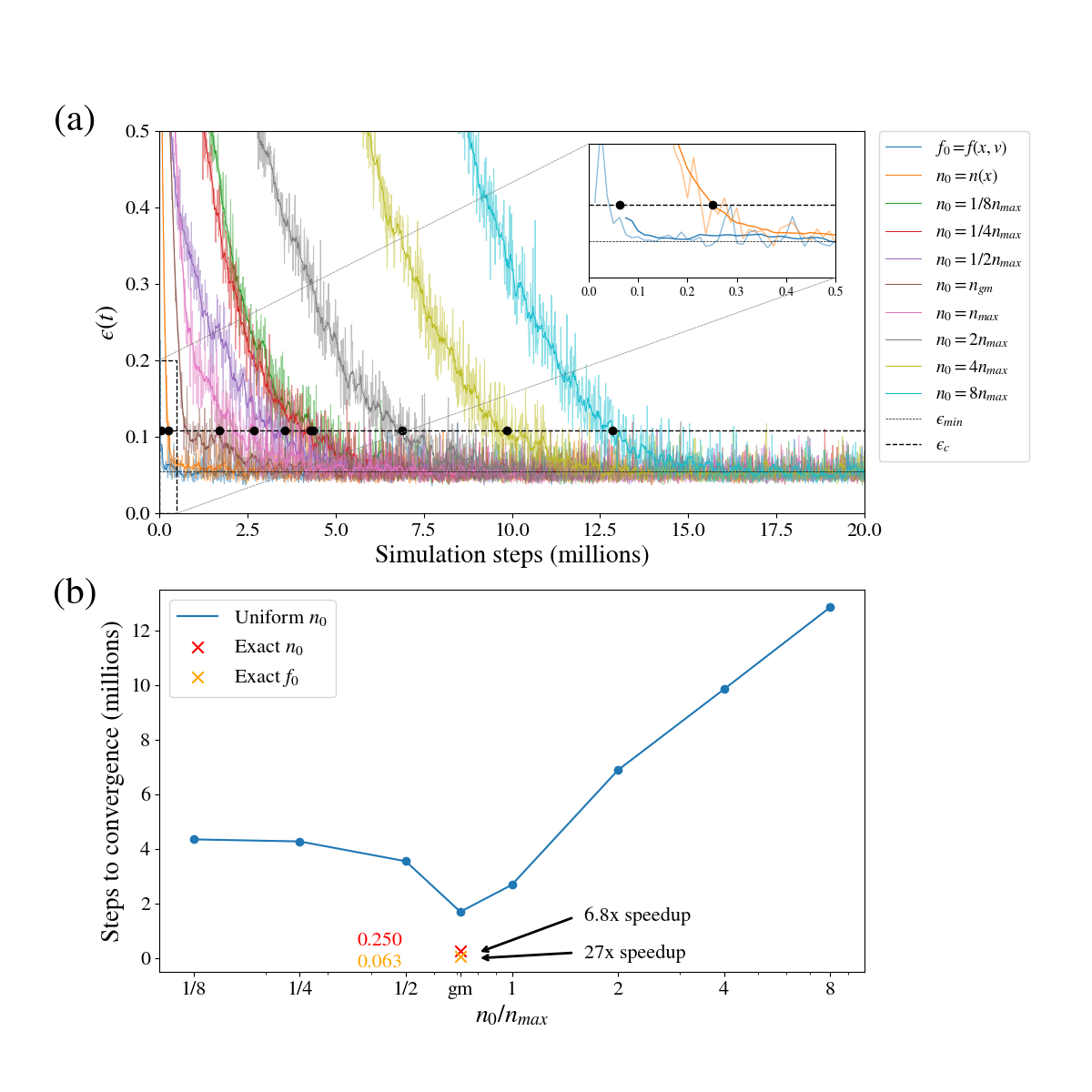}
    \caption{(a) Convergence error vs number of simulation time steps for various initial conditions (ICs). Black dots indicate the point where convergence is achieved for each simulation with respect to the \textit{offline} test procedure. (b) Steps to convergence for various ICs vs the ratio of initial density to the peak density of the steady-state. The red and orange crosses indicate the number of steps required when initializing the simulation with the exact final ion density profile $n$ or final IVDF $f$. The states speedup values for these crosses are with respect to blue dot above at $n_0/n_{max}=n_{gm}$.}
    \label{fig:converge_test}
\end{figure}

Figure \ref{fig:converge_test}a shows the evolution of the convergence error $\epsilon(t)$, as defined by Eq. \ref{eq:conv1}, vs time step for all ICs. The convergence time data is summarized in Fig. \ref{fig:converge_test}b with the x-axis showing the ratio of initial density to the peak ion density of the time-averaged profile $n_{max}$. The steps to convergence for the exact ion density and velocity distribution function initial conditions are placed under the $n_0/n_{max}=n_{gm}$ point.

Figure \ref{fig:converge_test} demonstrates that, for this single point in the parameter space, using the one-dimensional $n$ data as an IC offers a 6.8x speedup over even the best choice in uniform IC (when $n_0/n_{max}=n_{gm}$). Therefore higher order moments of the ion distribution function, as well as the quasi-steady electron distribution function reach convergence much faster than the ion density profile. Initializing a simulation with the two-dimensional $f$ offers even better performance with a 27x speedup in convergence, demonstrating that improved ICs can result in over an order of magnitude reduction in the number of time steps required to reach convergence.

The figure also reveals several other noteworthy trends. In particular, when initializing with a uniform ion density, fastest convergence is achieved when the initial value matches that predicted by the global model. Overestimating the initial density, i.e. choosing a value higher than the optimal value, quickly degrades convergence performance. The trend is approximately logarithmic, similar to that in the example Eq.'s \ref{eq:ex1}-\ref{eq:error_example} presented in Section \ref{sec:intro}. Despite this supposedly weak, logarithmic dependence on initial condition, we note that the number of steps required to reach convergence is still on the order of 10s of millions. From a physical standpoint, the observed trend is likely due to the fact that excess ions must be transported out of the system to achieve the correct final profile. However, selecting an initial uniform density below this peak results in only a modest increase in convergence time. In this case the system generates new ions near where additional density is required, minimizing the need for long-range ion transport. The takeaway is that when employing a uniform IC, a simple global model estimate might offer the best performance (see Appendix B for implementation details), and it is generally preferable to underestimate rather than overestimate the initial ion density.

The insert in Fig. \ref{fig:converge_test}a also shows that the simulation initialized with the true $f$ reaches convergence within the first tested time step (i.e. the width of the temporal averaging window defined in Section \ref{sec:offline_conv}). This suggests that a convergence test with higher temporal resolution may measure an even higher speedup in convergence time.

\subsection{Convergence time with MLP generated initial conditions for ion density}
\label{sec:fnn_results}

\begin{figure}[H]
    \centering
    \includegraphics[width=0.7\linewidth]{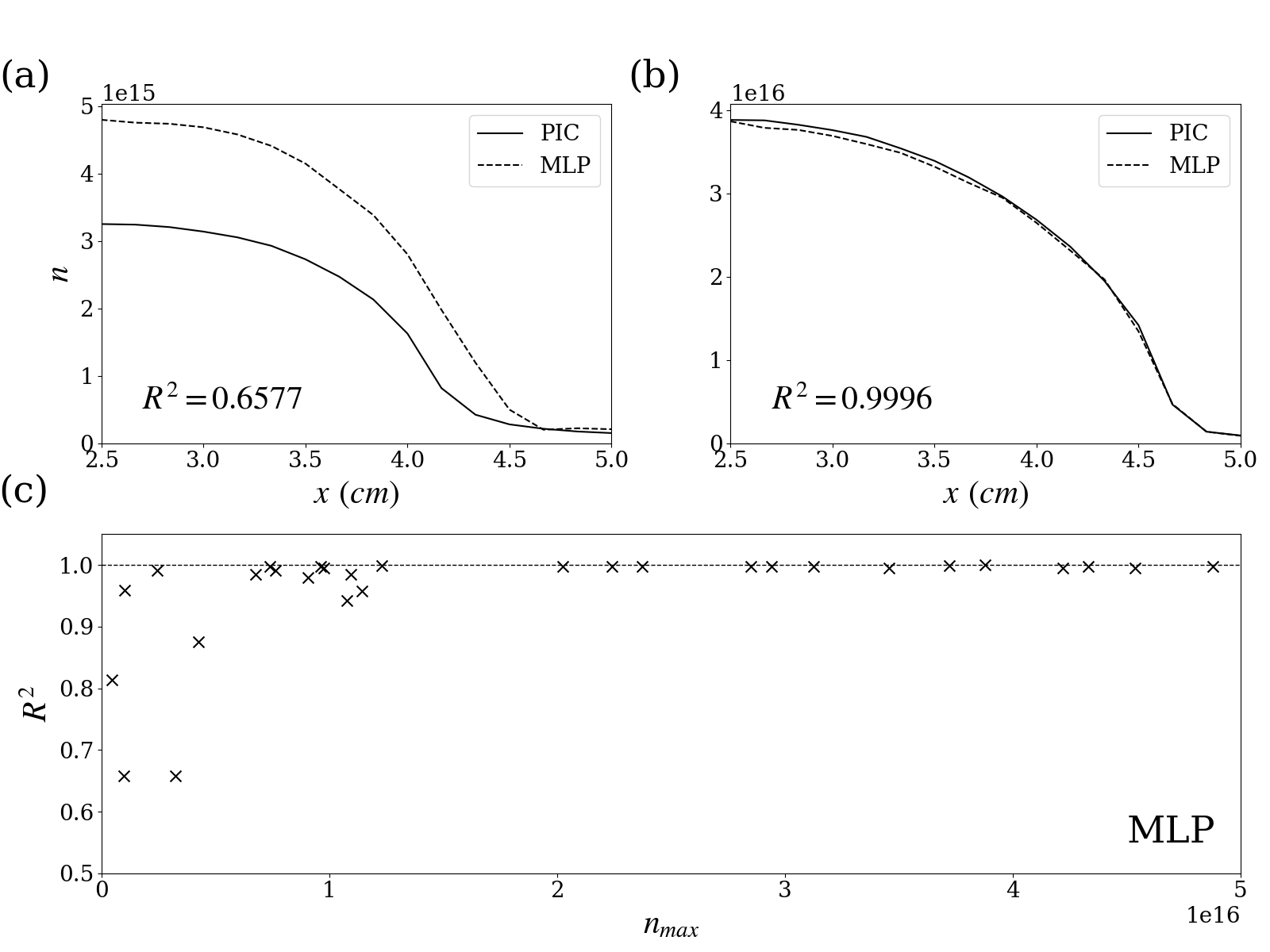}
    \caption{PIC simulation vs MLP generated time-averaged ion density profiles (on the half domain) for (a) the worst $R^2$ value and (b) the best $R^2$ value, and (c) plots of $R^2$ for the 40 test cases vs peak profile density ($n_{max}$) indicating that the MLP performs worst at generating low density profiles.}
    \label{fig:dnn_profiles}
\end{figure}

With confidence that our approach could yield convergence speedup, we proceed to test the MLP model described in Section \ref{sec:meth_dnn} as an initial condition generator (ICG) for $n$ profiles. Figure \ref{fig:dnn_profiles} shows several half $n$ profiles from the test set, including their true curves and those predicted by the MLP, these include the worst fit (Fig. \ref{fig:dnn_profiles}a) as well as the best fit (Fig. \ref{fig:dnn_profiles}b), which is more typical of most results. Figure \ref{fig:dnn_profiles}c plots the $R^2$ error of each test case against peak density, revealing that the model performed worst when predicting lower density (lower $n_{max}$) profiles. This is likely because absolute MSE, rather than normalized MSE was chosen for the loss function. Fortunately, simulations with lower density are both computationally cheaper, and converge more rapidly, even when using typical uniform ICs. Therefore, the model falls short within a parameter range where its purported improvements are least needed. Across the entire test set the model achieved a mean $R^2$ score of 0.958 and median of 0.996, demonstrating its ability to accurately generate $n$ profiles. 

When the 40 test cases were initialized with MLP generated ICs they on averaged converged, using an offline metric, \textbf{5.1x} faster than those initialized with a uniform density predicted from the global model. When using an online test for convergence this value decreased to \textbf{2.5x}. To determine the maximum theoretical speedup possible with this approach, each simulation was also initialized with the exact time-averaged quasi-steady-state ion density profile, providing a mean speedup of 6.5x (2.9x online). Therefore the MLP driven ML approach achieved 78\% (88\% online) of maximum possible reduction in simulation steps.

Translating these performance improvements into practical terms, the mean simulation time to reach convergence across the 40 test cases was 3.0 hours when measured with offline convergence (3.4 hours with online convergence), which was reduced to 0.59 (1.4) hours when initializing with the ICG. With respect to computing resources, the mean theoretical cost across the 40 test cases was 5.8 (6.6) petaFLOPs, which was reduced to 1.1 (2.6) petaFLOPs. In summary, these results demonstrate that even a relatively simple MLP can successfully accelerate computationally expensive kinetic plasma simulations of CCPs.

\subsection{Convergence time with PCA \& MLP generated initial conditions for IVDF}
\label{sec:pca_results}

\begin{figure}[H]
    \centering
    \includegraphics[width=0.7\linewidth]{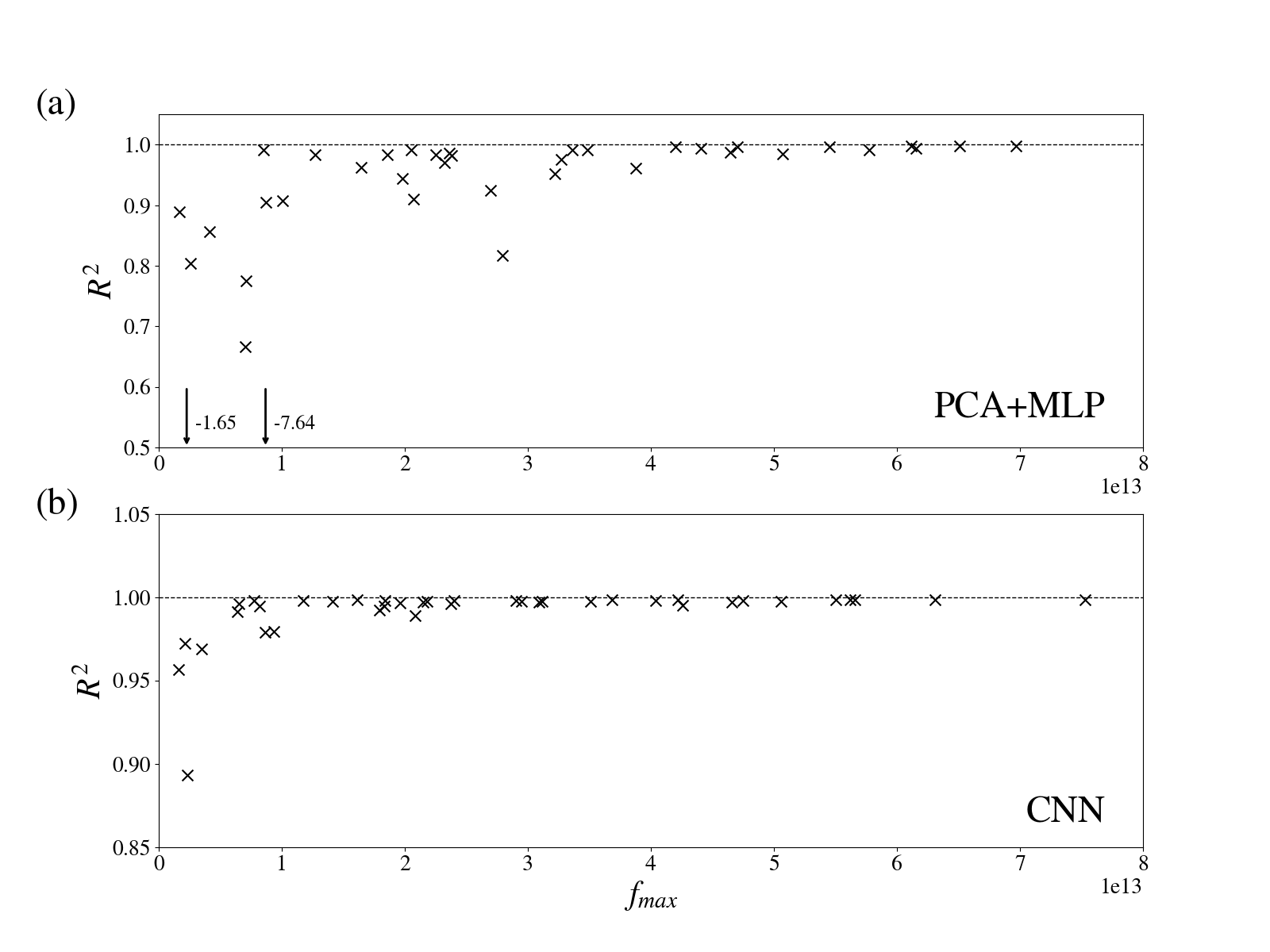}
    \caption{$R^2$ vs peak phase space density ($f_{max}$) for the 39 non-zero test cases for (a) the PCA+MLP model and (b) the CNN model. The position of outliers in (a) are denoted by the arrows and associated numerical values.}
    \label{fig:cnn_pca_profiles}
\end{figure}

Turning to predictions of the 1D-1V IVDF $f$ we first consider the performance of the PCA+MLP architecture, as described in Section \ref{sec:meth_pca_dnn}. Figure \ref{fig:cnn_pca_profiles}a shows that the model produces similar trends in $R^2$ error, with the worst performing predictions occurring for lowest maximum phase space density $f_{max}$. The mean $R^2$ error was 0.660, however the median was 0.983, which can be explained by the outliers observed in the figure at very low phase space densities. The model gave good predictions for most cases, with select visual comparisons provided for both 2D models and true $f$ profiles in Appendix C.

On average, test cases initialized with the PCA+MLP ICG converged, using an offline metric, \textbf{7.6x} faster than those with a uniform density initial condition predicted by the global model. When using an online test for convergence this value decreased to \textbf{2.9x}. To determine the maximum theoretical speedup possible with this approach, each simulation was also initialized with the exact time-averaged quasi-steady-state IVDF, providing a mean speedup of 19.5x (4.6x for online convergence). Therefore our ML driven approach achieved 39\% (63\%) of the maximum possible reduction in simulation steps. These numbers corresponded to a mean simulation time reduction from 3.0 (3.4) hours to 0.39 (1.2) hours when initializing with the ICG. The mean compute cost was reduced from 5.8 (6.6) petaFLOPs to 0.76 (2.8) petaFLOPs. These results demonstrate that a more complex, but still cheap to train, PCA+MLP model can further accelerate kinetic plasma simulations of CCPs.

\subsection{Convergence time with CNN generated initial conditions for IVDF}
\label{sec:cnn_results}

Finally we test the performance of the CNN model described in Section \ref{sec:meth_cnn} at predicting IVDF profiles. Figure \ref{fig:cnn_pca_profiles}b shows that the model produced similar trends of $R^2$ error to the other models, with the worst performing predictions occurring for lowest maximum phase space density $f_{max}$. Both the mean and median $R^2$ errors, 0.991 and 0.998 respectively, indicate that the model had the best predictive capabilities overall (see Appendix C for example generated $f$ profiles).

On average, test cases initialized with the CNN generated ICs converged, using an offline metric, \textbf{17.1x} faster than those using a uniform initial condition predicted by the model. When using an online test for convergence this value decreased to \textbf{4.4x}. Compared to the maximum available speedup using an IVDF initial conditions, our CNN ICG achieved 88\% (96\% online) of the maximum possible reduction in simulation steps. The mean simulation time was reduced from 3.0 (3.4) hours to 0.18 (0.78) hours when initializing with the ICG. The mean compute cost was reduced from 5.8 (6.6) petaFLOPs to 0.34 (1.5) petaFLOPs. These results demonstrate that the CNN model provided the best performance for accelerating kinetic plasma simulations of CCPs.

\subsection{Summary of speedup provided by ML generated initial conditions}
\label{sec:results_summary}

Table \ref{tab:icg_summary} summarizes the model accuracy and mean simulation speedup achieved by each of the simulation initialization approaches. A broad conclusion which can be drawn from these results is that machine learning generated initial conditions can accelerate the time to convergence for kinetic simulations of capacitively coupled plasma discharges by up to \textbf{17x}, when measured with the offline convergence criterion, and \textbf{4.4x} with the online convergence criterion. Generated profiles of ion density $n$ using MLPs can achieve near peak speedup when compared to initialization with the exact profiles. With respect to the 2D IVDF initial conditions, we see different performance levels between the two models. The CNN model performs best overall, reaching near peak speedup when compared to the exact IVDF profiles. The PCA+MLP shows reduced performance when compared to the CNN model, however still outperforms the MLP model for density alone. While the PCA+MLP model performed well at estimating most initial conditions, the outlier cases (see Fig. \ref{fig:cnn_pca_profiles}a) whose generated IVDF profiles led to longer convergence times, skewing the mean performance numbers. Improvements may be possible by increasing the threshold of explained variance, resulting in additional components in the PCA decomposition.

\begin{table}[H]
    \centering
    \caption{Summary of ICG performance for accelerating kinetic simulations of CCPs.}
    \label{tab:icg_summary}
    \begin{tabular}{r|c|c|c|c|c|c}
        \toprule
        Initial Condition & \makecell{No. sims\\training} & \makecell{No. sims\\testing} & Mean $R^2$ & Median $R^2$ & \makecell{Offline\\speedup} & \makecell{Online\\speedup} \\
        \midrule
        Uniform GM      & -  & 40 & - & - & 1.0 & 1.0 \\
        Exact $n$       & - & 40 & - & - & 6.1 & 2.9 \\
        MLP for $n$     & 210 & 40 & 0.958 & 0.996 & 5.1 & 2.5 \\
        \midrule
        Exact $f$       & - & 39 & - & - & 19.5 & 4.6 \\
        PCA+MLP for $f$ & 195 & 39 & 0.660 & 0.983 & 7.6 & 2.9 \\
        CNN for $f$     & 195 & 39 & 0.991 & 0.998 & 17.1 & 4.4 \\
        \bottomrule
    \end{tabular}
\end{table}

The results also clearly demonstrate that realizing these performance improvements in practice will be limited by the technique chosen for online testing of simulation convergence. Compared to the offline convergence criterion, the online approach leaves approximately 3x of the available performance improvement on the table. The largest differences in online and offline speedup occur for the IVDF initial conditions, primarily due to the very fast convergence times of simulations initialized with the PCA+MLP, and particularly, the CNN generated ICs. As shown within the insert axes of Fig. \ref{fig:converge_test}a, simulations can often converge quicker than the temporal averaging period used to check convergence, making the measured convergence time highly sensitive to the choice in averaging period, as well as the chosen waiting time to ensure that the error does not bounce back over the designated threshold.

To provide some qualitative examples of how the convergence criterion can influence measured speedup time we consider altering some of the parameters set for offline and online convergence, as defined in Section \ref{sec:convergence_def}. First, considering the offline convergence criterion, when reducing the averaging window for $\epsilon(t)$ from 10 to 5 periods, and increasing the convergence threshold from $2 \epsilon_{min}$ to $3 \epsilon_{min}$, the CNN ICG provides a 26x speedup over uniform initial conditions when compared against the global model (compared to 17.1x reported in Table \ref{tab:icg_summary}). Further decreasing the averaging window to just 1 period, and the convergence threshold to $4 \epsilon_{min}$, provides a measured speedup of 38x. Turning to the online convergence criterion, reducing the backwards averaging window of $\epsilon(t)$ from 10 to 5 periods, the convergence threshold from $2 \hat{\epsilon}_{min} (t)$ to $3 \hat{\epsilon}_{min} (t)$, and reducing the number of consecutive periods required for convergence from 25 to 10, we see a speedup of 6.4x (compared to 4.4x reported in Table \ref{tab:icg_summary}). Further reducing the averaging window to 1 period, and the required number of consecutive periods under the threshold to 5 (while leaving error levels unchanged), provides a measured speedup of 8.8x.

Of course choosing more relaxed definitions of convergence reduces the modelers certainty as to when convergence has actually been achieved, therefore the user should choose their criterion based on the given application and accuracy requirements. We further emphasize that future research should be directed into more efficient techniques to measuring PIC simulation convergence.

The machine learning techniques used in this work have proven to be effective at generating accurate ICs for PIC simulations, with the CNN model achieving 88\% and 96\% of maximum speedup for the offline and online convergence criterion respectively. When considering sparser datasets or larger parameter spaces, however, more advanced techniques may prove advantageous. Notable techniques include physics informed neural-networks, whereby physical laws or constraints are incorporated into the loss function \cite{raissi2019physics,li2025novel}, U-Nets, which offer improved multiscale representation \cite{wang2024automatically,van2025machine,khrabry2025hierarchical}, and diffusion models \cite{ramunno2024solar}, which have become the de facto technique for image generation (analogous to the $n$ and $f$ profiles considered here).

\section{Workflow for digital twin model generation}
\label{sec:digital_twin}

At this point, it would not be unreasonable for the reader to raise questions regarding the overall computational efficiency of the approach outlined thus far. Running 100s of simulations for the purpose of accelerating a few additional simulations is clearly a poor use of computational resources. However we reiterate that the purpose of this methodology is to accelerate convergence of costly first principles simulations for computer-aided engineering of complex devices, with the example here being a CCP discharge for silicon dry etching. The approach will prove computationally advantageous when a large number of simulations are required to explore a relatively well defined range of input parameters.

We also reemphasize that the ultimate goal of these simulations is not necessarily to obtain an ML model for the relatively low order behavior embedded in the ICs, but rather to accelerate solution to the more complex physics obtained from the high-fidelity simulations. For silicon etching this is the precise ion flux, energy and angular distributions at the interface between the plasma and the wafer. The relationship between the input parameters and these quantities are complex and highly non-linear, and would require significantly more data and/or more sophisticated techniques to generate a reduced order model (ROM) or digital (DT) twin which can capture these effects.

\begin{figure}[H]
    \centering
    \includegraphics[width=0.6\linewidth]{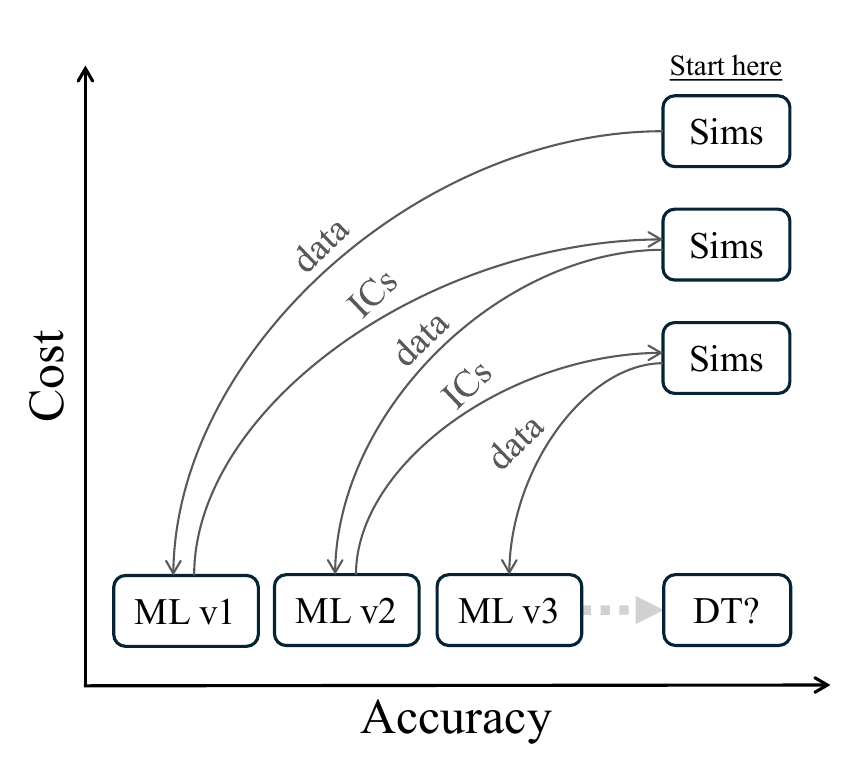}
    \caption{Workflow for reducing the cost of first principles simulations (designated as ``Sims''), while simultaneously improving the accuracy of an ML model for IC generation. Repetition of this cycle could help produce a reduced order model, or digital twin, with sufficient accuracy to enable design optimization or real time control.}
    \label{fig:dt_workflow}
\end{figure}

Figure \ref{fig:dt_workflow} demonstrates a workflow by which traditional first-principles physics simulations (designated as ``Sims'' in the figure) could be used in combination with ML techniques to accelerate collection of data for construction of a ROM or DT. The modeler first identifies an appropriate parameter space over which to train the model. In this paper that is a range of frequency $F$ and gas pressures $P$ inputs for a CCP discharge. The workflow begins by first exploring this parameter space using sparsely sampled PIC simulations. These are then used to train a first version of the ML model, which in turn accelerate the next round of PIC simulations applied to explore the parameter space with greater resolution. Iterating on this methodology will allow for progressively faster PIC simulations (shown as the Sim box moving down the ``Cost''-axis in Fig. \ref{fig:dt_workflow}) and improved ML model accuracy (shown as the ML box moving right on the ``Accuracy''-axis). Eventually, with sufficient data collection, the ML model may be sufficiently accurate to avoid running PIC simulations for many regions of the parameter space. Such a model is represented, aspirationally, by the ``DT'' box shown in the figure. This model could then be called on to rapidly prototype within the explored parameter space, or act as a digital twin for real time control and optimization of device performance. The high-fidelity PIC simulations would remain as a backup when additional resolution or certainty is needed.

An implied hypothesis in this procedure is that the ML model will enable faster simulations, even when the parameter space is sparsely sampled. To show that this is true for CCP simulations accelerated by initial conditions the CNN model described in Section \ref{sec:meth_cnn} was training on increasingly sparse datasets, and tested for its ability to accelerate convergence of the 39 (non-zero) test simulations, the results of which are shown in Table \ref{tab:dt_workflow}. It was observed, that even with an extremely sparse sampling of 3x3 samples across the parameter space, the model could realize a 2.7x speedup with offline convergence and 2.0x speedup with online convergence when compared to initialization with uniform density estimated by the global model. Moreover, speedup improvements continue until saturation at the levels reported in Section \ref{sec:cnn_results}. Estimating the compounding effects of this workflow, as outlined in Fig. \ref{fig:dt_workflow}, suggest that collecting data for the 195 (non-zero) simulations could be accelerated by a factor of \textbf{3.6x}, assuming online convergence testing, or \textbf{8.2x} if a convergence test approaching the offline speedup measures could be implemented. Table \ref{tab:dt_workflow} also reveals that our model saturates in performance after incorporating data from only 56 simulations, suggesting that fewer runs were required to realize the speedups reported in Section \ref{sec:acceleration}.

\begin{table}[H]
\centering
\caption{CNN accuracy and PIC simulation speedup achieved at various iterations of the digital twin training workflow shown in Fig. \ref{fig:dt_workflow}. In this instance the model architecture remains fixed and each version of the model simply incorporates additional data.}
\label{tab:dt_workflow}
\begin{tabular}{c|c|c|c|c|c}
  \toprule
  Iteration & \makecell{No. sims\\training} & Mean $R^2$ & Median $R^2$ & \makecell{Offline\\speedup} & \makecell{Online\\speedup} \\
  \midrule
  I   & 9   & 0.802 & 0.942 & 2.7 & 2.0 \\
  II  & 16  & 0.940 & 0.992 & 7.2 & 3.4 \\
  III & 25  & 0.925 & 0.995 & 10.5 & 3.9 \\
  IV  & 42  & 0.943 & 0.996 & 12.8 & 4.0 \\
  V   & 56  & 0.967 & 0.997 & 16.9 & 4.4 \\
  VI  & 195 & 0.991 & 0.998 & 17.1 & 4.4 \\
  \bottomrule
\end{tabular}
\end{table}

There are some important challenges to consider when deploying this procedure to a larger, or higher dimensional, parameter space. For one, the hypothesis regarding monotonic improvements in ML performance may not hold true for very sparse samples over a large parameter space, potentially requiring a larger upfront cost in initial, full cost, simulations prior to beginning the procedure outlined in Fig. \ref{fig:dt_workflow}. Alternatively, the version numbers associated with each ML model in the figure need not only refer to model improvement associated with training on additional data, but could also refer to different model architectures or machine learning techniques entirely (see the final paragraph of Section \ref{sec:results_summary} for examples). The ML model chosen for each level should be optimized to take advantage of the amount of available data.

Another consideration is how best to sample from the a larger or higher dimensional parameter space. Simulations in this work were sampled nearly uniformly across the two-dimensional space, an approach which could prove sub-optimal in higher dimensions. Random sampling, Latin hypercube sampling \cite{McKay1979LHS,Stein1987LHS} or space filling curves \cite{bader2012space} offer viable approaches. Furthermore if the ML method can be modified to provide uncertainty quantification, Bayesian optimization \cite{shahriari2015taking} techniques may help inform the sample locations which best improve model performance. A suitable stopping criterion for data collection could be met once PIC simulation speedup has saturated, or when an ML model for the desired output (such as surface IVDFs) has reached a sufficient accuracy. Exploring these ideas will be the subject of future research.

\section{Conclusion}
\label{sec:conclusion}

This paper has demonstrated that machine learning techniques, specifically multi-layer perceptrons, optionally with principal component analysis, and convolutional neural network, can be applied to develop an accurate mapping between the operating parameter space of capacitively coupled plasma discharges and the final time-averaged ion density and velocity distribution profiles. These profiles, can in-turn, be used to accelerate further 1D-3V kinetic simulations by up to a factor of 17.1x, with an offline measure of convergence, or 4.4x, when using an online measure of convergence. This approach will prove most useful when deep exploration of a given parameter space is required, as is common for computer aided engineering design. By reducing the computational cost associated with collecting further, a natural workflow emerges to assist in the development of robust reduced order models, or eventually, digital twins which could be used for online optimization and control. 

This work offers many opportunities for extension, including exploration of a wider parameter range, learning of higher fidelity initial conditions (perhaps with temporal characteristics of the distribution functions), and applying the 1D(-1V) results to generate initial conditions for more costly higher dimensional simulations. The procedures demonstrated here can also likely be applied to other multi-time-scale plasma systems, providing improvements in simulation performance and reduced time for design of new technologies.


\section*{Acknowledgments}

This work was supported by the AI/ML program of the U.S. Department of Energy (DOE) Office of Fusion Energy Science (FES) through grant number DE-SC0024522.

The simulations presented in this article were performed on computational resources managed and supported by Princeton Research Computing, a consortium of groups including the Princeton Institute for Computational Science and Engineering (PICSciE) and the Office of Information Technology's High Performance Computing Center and Visualization Laboratory at Princeton University.

This research used resources of the National Energy Research Scientific Computing Center (NERSC), a U.S. Department of Energy Office of Science User Facility located at Lawrence Berkeley National Laboratory, operated under Contract No. DE-AC02-05CH11231 using NERSC Award No. BES-ERCAP0032989.

\section*{Data availability}

The data that support the findings of this study are available from the corresponding author upon reasonable request.

\section*{Appendix A: Why GPUs are a bottleneck for faster time-steps}
\label{sec:gpu_bottleneck}

Although the grid and particles within a PIC simulation are well-suited for parallelization and hardware acceleration, the tightly coupled nature of the equations means that time-stepping is inherently serial. If each time-step of the simulation takes up a fixed computer wall-clock-time, then this sets a lower bound on the total simulation runtime. In this paper we explored the use of an initial condition generator (ICG) to reduce the number of time steps to reach convergence, however the reader could just as well ask. Why don't we reduce the wall-clock-time per step? Interestingly, the increasing reliance on modern accelerators such as GPUs tends to constrain the potential gains available through this approach.

\begin{figure}[H]
    \centering
    \includegraphics[width=0.6\linewidth]{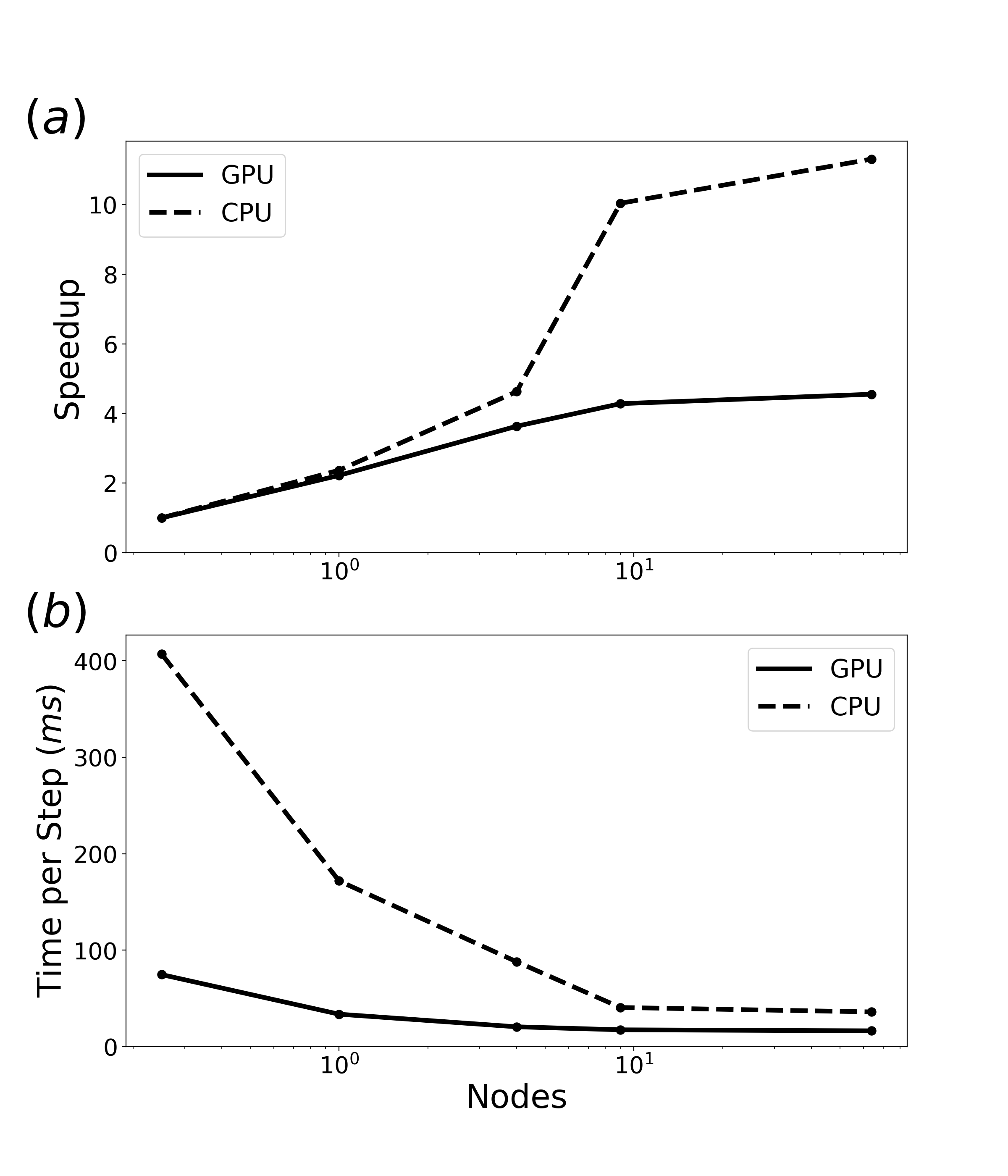}
    \caption{Strong scaling performance of the LTP-PIC code \cite{powis2021particle} when modeling a two-species thermal plasma with all-periodic boundary conditions. The grid size is $256 \times 256$ and there are approximately 130 million simulation particles. (a) Speedup plot demonstrating that CPUs offer improved strong scaling over GPUs, and (b) plot of time per step demonstrating that despite scaling limitation GPUs still offer improved performance. Each node was comprised of 1x AMD EPYC 7763 CPU with 64 cores and 4x NVIDIA A100 GPUs. The leftmost data point was obtained using 1/4 of a node (16 CPU cores or 1 GPU).}
    \label{fig:weak_scaling}
\end{figure}

When adapting an algorithm for execution on a GPU, the prevailing best practice is to expose as much parallelism and data throughput as possible, thereby leveraging the chip’s massively parallel architecture. For suitably structured algorithms, this yields significant performance improvements, even though individual GPU cores operate at lower clock speeds compared to typical CPU cores. However, this architectural design comes with limitations—specifically, diminished strong scaling efficiency relative to CPUs (see Fig. \ref{fig:weak_scaling}a), which in turn restricts the performance benefits achievable through increased resource allocation. Nonetheless, as illustrated in Fig. \ref{fig:weak_scaling}b, GPUs consistently outperform CPUs across all node counts, demonstrating a clear net speedup despite their scaling constraints.

The intent of this discussion is to highlight the limitations users face in further accelerating individual time steps on modern HPC architectures. While GPU-based nodes remain the most effective option for minimizing total wall-clock time (not to mention energy usage), their limited strong scaling capabilities restrict the extent to which performance can be improved through additional resource allocation. This limitation motivated the approach pursued in this work, which rather aims to reduce the number of time steps required to reach a desired level of accuracy or convergence.

\section*{Appendix B: Inhomogeneous argon capacitively coupled plasma global model}
\label{sec:global_model}

The global model used to predict plasma density in Section \ref{sec:acceleration} is an implementation of the inhomogeneous global model for argon CCPs from Chabert \cite{chabert2011physics} Chapter 5, which draws directly from the work of Godyak \cite{godyak1986soviet} and Lieberman \& Lichtenberg \cite{lieberman1994principles}. The model is derived under the following assumptions:

\begin{enumerate}
    \item The electron temperature $T_e$ (given in units of $eV$) is constant in space.
    \item Ion inertia is sufficient large such that ions respond only to the time-averaged electric field.
    \item Electron inertia is negligible. Electrons therefore follow the instantaneous electric field.
    \item The system is electrostatic.
    \item Gas pressure is constant in space and within the low to intermediate range (under 100 mTorr).
    \item There is no volume recombination or multi-step ionization.
    \item The sheath size is small compared to the discharge length.
\end{enumerate}

\subsection*{Particle balance}

At lower pressures, the steady state particle balance can be written as,

\begin{equation}
    n_g K_{iz} (l - s_m) = 2 h_l u_B,
\end{equation}
where $n_g$ is the gas density, $l$ is the discharge length, $s_m$ is the sheath length and $K_{iz}$ is the ionization rate coefficient defined by,

\begin{equation}
    K_{iz} = K_{iz0} \exp{\left( -\frac{\varepsilon_{iz}}{T_e} \right)},
\end{equation}
where $K_{iz0}=5.0\times10^{-14}$ $1/m^3s$ 
and $\varepsilon_{iz}=17.44$ $eV$ is the argon ionization potential. 
$h_l$ is the edge to center density ratio, which in the low to intermediate pressure regime is given by,

\begin{equation}
    h_1 \approx 0.86 \left[3 + \frac{l}{2 \lambda_i}  \right]^{-1/2},
\end{equation}
where $\lambda_i$ is the ion-neutral mean-free-path, which for argon can be approximated as,

\begin{equation}
    \lambda_i = \frac{31.5 \times 10^{-3}}{P},
\end{equation}
where $P$ is the gas pressure in units of $mTorr$. 
$u_B$ is the Bohm speed,

\begin{equation}
    u_B = \left( \frac{e T_e}{M} \right)^{1/2},
\end{equation}
where $M$ is the ion mass.

When solving the global model we consider the particle balance equation in the following form,




\begin{equation}
\label{eq:particle}
\begin{split}
    f(T_e,s_m) &= \frac{n_g(l - s_m)}{2 h_l} - \frac{u_B}{K_{iz}} \\
    &= \frac{n_g K_{iz0} (l - s_m)}{2 h_l} - \left( \frac{e T_e}{M} \right)^{1/2}  \exp{\left(\frac{\varepsilon_{iz}}{T_e} \right)} 
\end{split}
\end{equation}

\subsection*{Power balance}

The steady-state electron power balance can be written via a circuit representation. Since our system is voltage driven, we consider the power balance in terms of voltage terms,

\begin{equation}
\label{global_power}
    \frac{1}{2} \left( V_{ohm} + 2 V_{stoc} + 2 V_{ohm,sh} \right) J_0 = 2e  h_l n_0 u_B \varepsilon_T(T_e),
\end{equation}
where $V_{ohm}$ is given by,

\begin{equation}
    V_{ohm} = K_{ohm} h_l m \nu_m (l - 2 s_m) \left( \frac{\omega}{e} \right)^{3/2} \left( \frac{\varepsilon_0 s_m e T_e}{J_0} \right)^{1/2}.
\end{equation}

$J_0$ is the current per unit area of electrode (since our problem is 1D) and $n_0$ is the peak plasma density. Then we have $K_{ohm} = 1.55$ in a collisionless (low-pressure) sheath 
and $K_{ohm} = 1.14 \sqrt{s_m / \lambda_i}$ in a collisional (intermediate pressure) sheath. 
$\nu_m = n_g \bar{\sigma}_{el} \bar{v}_e$ is the electron-neutral elastic collision frequency, 
with $\bar{\sigma}_{el} = 1.0 \times 10^{-19}$ 
and $\bar{v}_e = (8 e T_e / \pi m)^{1/2}$. 
Next we have,

\begin{equation}
    V_{stoc} = K_{stoc} (meT_e)^{1/2} \left( \frac{\omega s_m}{e} \right),
\end{equation}
where $K_{stoc} = 0.72$ in a collisionless sheath 
and $K_{stoc} = 0.8$ in a collisional sheath. 
Next,

\begin{equation}
    V_{ohm,sh} = K_{ohm,sh} m \nu_m s_m \left( \frac{\omega s_m}{e} \right),
\end{equation}
where $K_{ohm,sh} = 0.33$ in a collisionless sheath 
and $K_{ohm,sh} = 0.155$ in a collisional sheath. 
$\omega \equiv \omega_{RF}$ is the driving angular frequency.

As for the right hand side of Eq. \ref{global_power} we have,
\begin{equation}
    \varepsilon_T(T_e) = \varepsilon_{iz} + \frac{K_{exc}}{K_{iz}} \varepsilon_{exc} + \frac{3m}{M} \frac{K_{el}}{K_{iz}} T_e + 2 T_e + \Delta \phi,
\end{equation}
where,
\begin{equation}
    K_{exc} = K_{exc0} \exp{\left( - \frac{\varepsilon_{exc}}{T_e} \right)},
\end{equation}
with $K_{exc0} = 0.16 \times 10^{-18}$ $1/m^3 s$ 
and $\varepsilon_{exc} = 12.38$ $eV$. 
We also have $K_{el} = \nu_m / n_g$, 
and finally, $\Delta \phi = 3/8 V_0$, 
(although $\Delta \phi = 1/2 V_0$ is a viable choice).

Writing the power balance as $n_0 = g(T_e,s_m)$ gives,

\begin{equation}
    \label{eq:power}
    n_0(T_e,s_m) = \frac{\left( V_{ohm}(T_e,s_m) + 2 V_{stoc}(T_e,s_m) + 2 V_{ohm,sh}(T_e,s_m) \right) J_0}{4 e h_l u_B(T_e) \varepsilon_T(T_e)}
\end{equation}

\subsection*{Sheath size}

The relationship between the sheath size $s_m$ and peak plasma density $n_0$ depends on whether the sheath is collisionless or weakly collisional. For the collisionless case we have,

\begin{equation}
    s_m = \frac{5}{12(e h_l n_0)^2 \varepsilon_0 T_e} \left( \frac{J_0}{\omega} \right)^3,
\end{equation}

In the collisional case, we have,

\begin{equation}
    s_m = 0.88 \left( \frac{\lambda_i}{\varepsilon_0 e^2 T_e \omega^3 h_l^2} \right)^{1/2} \frac{J_0^{3/2}}{n_0}.
\end{equation}

These are re-written as,

\begin{equation}
     \label{eq:sheath_nocoll}
    h(T_e,n_0,s_m) = s_m - \frac{5}{12(e h_l n_0)^2 \varepsilon_0 T_e} \left( \frac{J_0}{\omega} \right)^3,
\end{equation}

and,

\begin{equation}
     \label{eq:sheath_coll}
    h(T_e,n_0,s_m) = s_m - 0.88 \left( \frac{\lambda_i}{(e h_l n_0)^2 \varepsilon_0 T_e} \right)^{1/2} \left( \frac{J_0}{\omega} \right)^{3/2}.
\end{equation}





\subsection*{Relating current to voltage amplitude}

We will also need,
\begin{equation}
    V_0 \approx \frac{s_m J_0}{K_{cap} \omega \varepsilon_0},
\end{equation}
where $K_{cap} = 0.613$ in a collisionless sheath 
and $0.751$ in a collisional sheath. 

This gives,

\begin{equation}
    \label{eq:current_voltage}
    J_0 \approx \frac{K_{cap} \omega \varepsilon_0 V_0}{s_m}
\end{equation}

\subsection*{Solution algorithm}

The system of equations, \ref{eq:particle}, \ref{eq:power} and \ref{eq:sheath_nocoll} (or \ref{eq:sheath_coll}) with \ref{eq:current_voltage} are solved iteratively via a bisection approach. This is chosen due to the presence of small gradients in many of the functions. In the outermost loop we iterate over Eq. \ref{eq:sheath_nocoll} or \ref{eq:sheath_coll} for the sheath size. At each iteration this requires an internal iteration over the particle balance equation \ref{eq:particle} to compute the electron temperature. The power balance equation \ref{eq:power} is used directly to predict $n_0$.

To decide whether to use the collisional (intermediate pressure) or collisionless (low pressure) model, we compute the discharge properties with both models and choose that which best matches the true value from PIC simulations. This approach naturally provides more more conservative estimates for the results in this paper.

\section*{Appendix C: Select CCP IVDF profiles generated from kinetic PIC simulations and ML models.}

\begin{figure}[H]
    \centering
    \includegraphics[width=0.95\linewidth]{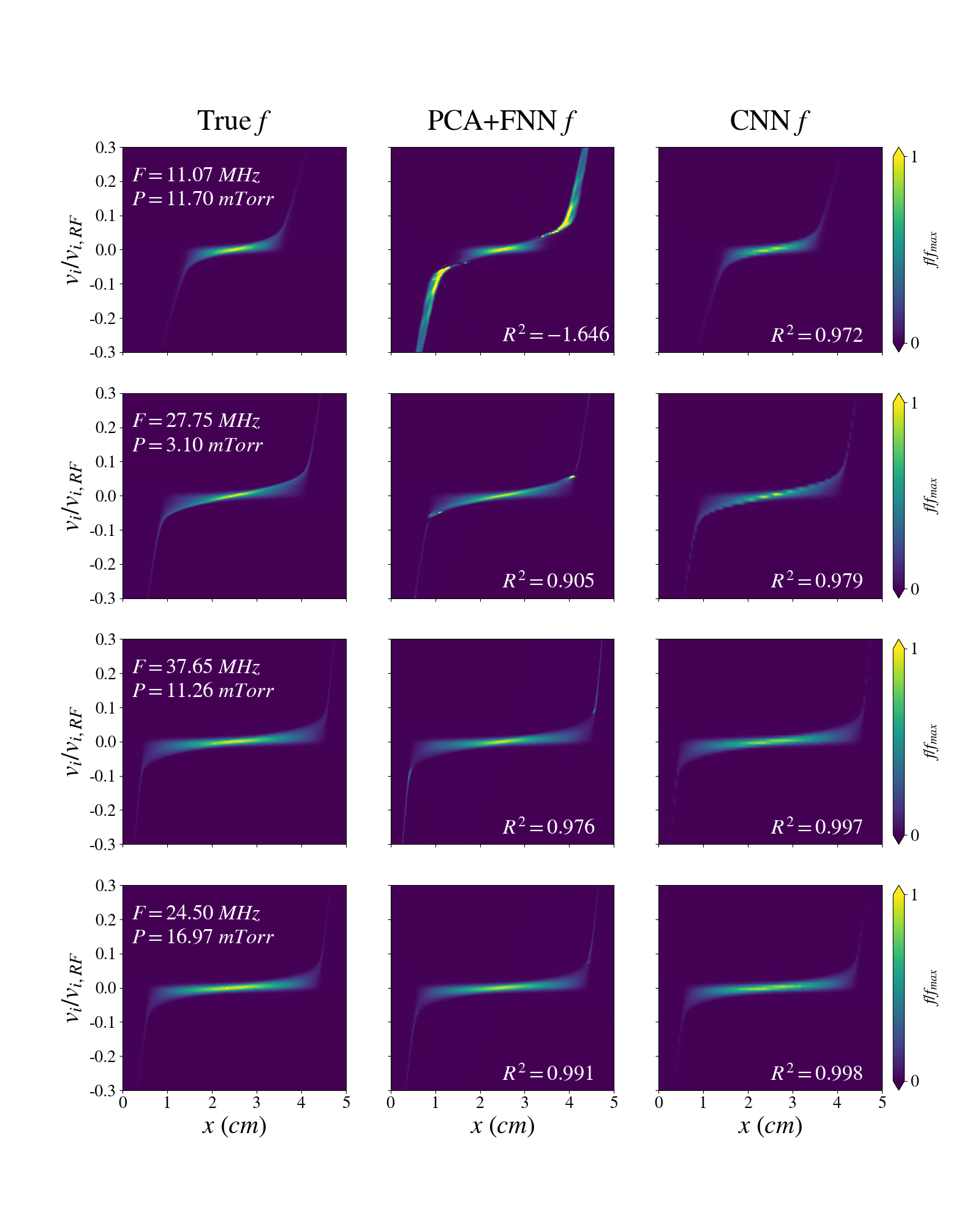}
    \caption{Selected 1D-1V IVDF profiles at four different CCP operating conditions. Each row represents a different operating condition with the gas pressure $P$ and driving frequency $F$ given at the top left of the figures in the left column. The left column represents the ``true'' IVDFs from time-averaged diagnostics of the PIC simulation after reaching quasi-steady state. The center column shows the IVDFs generated by the PCA+MLP model. The right column shows the IVDFs generated using the CNN model. $R^2$ values for each model are shown at the bottom right within their respective figures.}
    \label{fig:ex_ivdfs}
\end{figure}

\bibliographystyle{ieeetr}
\bibliography{sample}

\end{document}